\documentclass[twocolumn,preprintnumbers,superscriptaddress,prl,nofootinbib]{revtex4}
\usepackage{graphicx}
\usepackage{times}
\usepackage{epsfig}
\usepackage{amsmath}
\usepackage{amsfonts}
\usepackage{amssymb}
\usepackage{url}
\usepackage{subfigure}
\newcommand{\be}{\begin{equation}}
\newcommand{\ee}{\end{equation}}
\newcommand{\bea}{\begin{eqnarray}}
\newcommand{\eea}{\end{eqnarray}}

\begin{document}
\pagestyle{plain}
\title{
Gravity Waves and Gravitino Dark Matter 
in $\mu$-Hybrid Inflation
}
\author{Nobuchika Okada}
\affiliation{
Department of Physics and Astronomy,
University of Alabama,
Tuscaloosa, AL 35487, USA
}
\author{Qaisar Shafi}
\affiliation{
Bartol Research Institute,
Department of Physics and Astronomy,
University of Delaware, Newark, DE 19716, USA
}


\begin{abstract}

We propose a novel reformulation of supersymmetric (more precisely $\mu$-) hybrid inflation based 
  on a local U(1) or any suitable extension 
  of the minimal supersymmetric standard model (MSSM) which also resolves the $\mu$ problem. 
We employ a suitable Kahler potential which effectively yields quartic inflation with non-minimal coupling to gravity. 
Imposing the gravitino Big Bang Nucleosynthesis (BBN) constraint on the reheat temperature ($T_r \lesssim 10^6$ GeV) 
  and requiring a neutralino LSP, the tensor to scalar ratio ($r$) has a lower bound $r \gtrsim  0.004$. 
The U(1) symmetry breaking scale lies between $10^8$ and $10^{12}$ GeV. 
We also discuss a scenario with gravitino dark matter whose mass is a few GeV.

\end{abstract}
\maketitle
Minimal supersymmetric hybrid inflation \cite{ref1, ref2} is based on a unique renormalizable superpotential $W$, 
  employs a canonical Kahler potential $K$, and it is readily implemented with a local U(1) or other suitable extension 
  of the minimal supersymmetric standard model (MSSM). 
With only the tree level and radiative corrections included in the inflationary potential, the scalar spectral index $n_s$ lies close to $0.98$ 
  and the symmetry breaking scale is of the order of $10^{15}$ GeV \cite{ref1,ref3}. 
The inclusion \cite{ref4,ref5,ref6} of soft supersymmetry breaking terms plays an essential role in reducing 
  $n_s$ to the desired value $0.9655 \pm 0.0062$ measured by the Planck \cite{ref7} and the WMAP \cite{ref8} satellite experiments. 

An important feature of this class of supersymmetric inflation models is their ability to resolve the MSSM $\mu$ problem, 
  noted several years ago by Dvali, Lazarides and Shafi \cite{ref9}. 
The minimal model employs a U(1)$_ R$ symmetry which forbids the standard MSSM $\mu$ term and 
  also requires the presence of a gauge singlet chiral superfield $S$ which couples to the MSSM Higgs doublets. 
After supersymmetry breaking, the scalar component of $S$ (inflaton field in the standard scenario) acquires a non-zero 
  vacuum expectation value (VEV) which helps induce the desired MSSM $\mu$ term.

Inflation in this framework, dubbed ``$\mu$-hybrid inflation,'' has been studied in Refs.~\cite{OS} and \cite{RSV}. 
Ref.~\cite{OS} employs minimal $W$ and $K$ and concludes that such a framework yields a relatively high reheat temperature 
  $T_r \simeq 10^{11}$ GeV. 
It was concluded that this scenario is compatible with the so-called split supersymmetry \cite{ref12}. 
Namely, the gaugino partners of the SM gauge bosons lie in the presumably observable TeV range, 
  while all scalar particles except for the SM Higgs boson are much heavier ($\sim 10^ 7$ GeV or so). 
In Ref.~\cite{RSV}, however, the Kahler potential includes higher order terms which allows one to bring $T_r$ 
  down to $10^6-10^7$ GeV which is compatible with TeV scale supersymmetry and the BBN gravitino constraints.

In this paper we propose a novel reformulation of $\mu$-hybrid inflation based on the unique renormalizable superpotential 
  which also resolves the MSSM $\mu$ problem \cite{ref9}. 
The Kahler potential is suitably modified in order to implement a supersymmetric version of $\lambda \varphi^4$ inflation 
  with non-minimal coupling of $\varphi$ to gravity \cite{KLVP} 
  (see, also, Refs.~\cite{Arai:2011nq, Pallis:2011gr, Leontaris:2016jty} for applications to inflation scenarios 
   in supersymmetric Grand Unified Theories (GUTs)). 
In our scenario, instead of $S$, the scalar field that breaks the local U(1) or some other suitable symmetry plays the role of $\varphi$ 
  and therefore drives inflation. 
This enables us to keep the reheat temperature after inflation below $10^6-10^7$ GeV and therefore TeV scale supersymmetry 
  is compatible with the inflationary scenario. 
The symmetry breaking scale associated with inflation can lie in a rather wide range from $10^6$ GeV to the GUT scale, $M_{GUT} \simeq 2 \times 10^{16}$ GeV. 
We also consider a scenario with gravitino as dark matter with mass of order of 1 GeV.

In minimal supersymmetric hybrid inflation the renormalizable superpotential is expressed as
\begin{equation}
W = S \left[ \kappa(\bar{X}X-M^2) \right], 
\label{eq:1}
\end{equation}
where $X$, $\bar{X}$ denote a pair of conjugate superfields under the symmetry assumed for simplicity to be U(1), 
  $S$ is a singlet under this symmetry, $M$ denotes the symmetry breaking scale, and $\kappa$ denotes 
  a real dimensionless coupling parameter.
A U(1)$_R$ symmetry is assumed under which both $W$ and $S$ carry two units of charge 
  and the combination $\bar{X}X$ is invariant. 
Note that this symmetry prevents the appearance in $W$ of terms such as $S^2$ and $S^3$ that can ruin inflation.

In the supersymmetric vacuum, the underlying symmetry is broken and the scalar component of $S$, denoted with the same symbol, 
  has zero VEV. 
During inflation the scalar $S$ slowly rolls from sub-Planckian values towards $M$ and finally to the origin 
  (actually to a location displaced from the origin by an amount proportional to $m_{3/2}$, the gravitino mass.) 
A scalar spectral index $n_s$ lying between $0.96$ and $0.97$ is realized for $\kappa \simeq 10^{-3}$ and $M \simeq 10^{15}$ GeV. 
The tensor to scalar ratio $r$ is tiny but may approach $10^{-4}$ in some cases.

In $\mu$-hybrid inflation \cite{ref9} the following additional term is included in $W$:
\begin{equation}
\sigma H_u H_d S, 
\label{eq:2}
\end{equation}
where the dimensionless constant $\sigma > \kappa$, in order to prevent the MSSM doublets from acquiring VEVs of order $M$. 
Inflation proceeds as in the previous case except that reheating after inflation is determined by the $\sigma$ term above. 
A careful evaluation reveals that the reheat temperature exceeds $10^{11}$ GeV and the model is compatible with split supersymmetry \cite{OS}. 
As previously mentioned this conclusion can be circumvented by including higher order terms in the Kahler potential \cite{RSV}. 
This allows one to work with smaller $\sigma$ values which, in turn, yield suitably lower values for the reheat temperature 
  that are compatible with TeV scale supersymmetry.

In our reformulation of $\mu$-hybrid inflation we propose to modify the Kahler potential in a way that allows us 
  to implement quartic inflation with non-minimal coupling to gravity \cite{non-minimal}. 
A suitable combination of $X$, $\bar{X}$ fields, and not the field $S$, drives inflation in this scenario. 
The inflaton has trans-Planckian values during the slow roll epoch and eventually reaches its present day minimum value $M$. 
As we shall see, $M$ can take values anywhere from $10^6$ GeV to $10^{13}$ GeV, if the reheat temperature $T_r \lesssim 10^6$ GeV. 
Somewhat larger $T_r$ values, say of order $3 \times 10^7$ GeV, will allow $M$ to reach the GUT scale. 
The model also predicts observable gravity waves with $r \gtrsim 0.004$.

In order to implement non-minimal quartic inflation, we employ a Lagrangian with a non-minimal Kahler potential 
 in the superconformal framework of supergravity \cite{SUGRA}:
\begin{equation}
\mathcal{L} \supset \int d^4 \theta \; \phi^\dagger \phi \; (-3  M_P^2  \Phi), 
\label{eq:3}
\end{equation}
where $M_P=2.44 \times 10^{18}$ GeV is the reduced Planck mass, 
\begin{eqnarray}
  \Phi &=& 1 - \frac{1}{3 M_P^2} \left( |\bar{X}|^2+|X|^2+|S|^2 \right) \nonumber\\
    &&+\frac{1}{3 M_P^2} \gamma  \left(\bar{X}X+ (\bar{X}X)^\dagger \right) + \cdots, 
\end{eqnarray}  
   and $\phi=1+\theta^2 F_{\phi}$  is the compensating multiplet with 
   a supersymmetry breaking $\langle F_{\phi} \rangle $ = $m_{3/2}$ 
   with $m_{3/2}$ being gravitino mass.  
Note that $\Phi$ may include higher order terms for $S$ (denoted by $+\cdots$) 
   to stabilize the scalar potential in the $S$-direction \cite{KLVP}.

The relevant terms in the Jordan frame Lagrangian for inflation are as follows:
\begin{eqnarray}
    {\cal L} &\supset&  - \frac{1}{2} M_P^2 \Phi {\cal R} 
     +  \left(\partial_\mu \bar{X} \right)^\dagger \left(\partial^\mu \bar{X} \right) 
     +  \left(\partial_\mu X \right)^\dagger \left(\partial^\mu X \right) \nonumber \\
    && +  \left(\partial_\mu S \right)^\dagger \left(\partial^\mu S \right)
     - V_{SUSY}, 
\end{eqnarray}
where 
\begin{eqnarray}
   V_{SUSY} \supset \kappa^2 \left| \bar{X} X -M^2 \right|^2 + \kappa^2 \left| S \right|^2 \left( |{\bar X}|^2 + |X|^2\right). 
\end{eqnarray}
Now we consider the inflation trajectory along the $D$-flat direction, $\langle \bar{X} \rangle = \langle X \rangle$, 
   with the inflaton parametrized as $\bar{X}= X = (1/2) \varphi$.\footnote{
The mass of the fluctuations in the direction orthogonal to the $D$-flat direction 
   is estimated to be $\sim g \varphi $, where $g$ is the U(1) gauge coupling. 
For a sizable gauge coupling value, the scalar potential is tightly bounded in this orthogonal direction. 
Hence, it is justified to choose the $D$-flat direction as the inflation trajectory and parametrize it  
   by only one field $\varphi$. 
}    
During inflation the scalars $S$, $H_u$ and $H_d$ are at their potential minimum. 
Along the inflation trajectory for $\varphi \gg M$, the relevant Langrangian is simplified to be 
\begin{equation}
{\cal L} \supset -\frac{1}{2} M_P^2 (1+\xi \mathcal{\varphi}^2) {\cal R} 
  + \frac{1}{2} g^{\mu \nu}(\partial_\mu \varphi)(\partial_\nu \varphi)-\frac{\kappa^2}{16}\varphi^4, 
\label{Leff}  
\end{equation}
where the dimensionless parameter $\xi$ is given by
\begin{equation}
\xi = \frac{1}{6}(\gamma - 1).
\end{equation}
Note that Eq.~(\ref{Leff}) is nothing but the Jordan frame Lagrangian of  
   $\lambda \varphi^4$ inflation with non-minimal gravitational coupling 
   (see, for example, Ref.~\cite{ref13}).

\begin{figure}[t]
  \begin{center}
   \includegraphics[width=8cm]{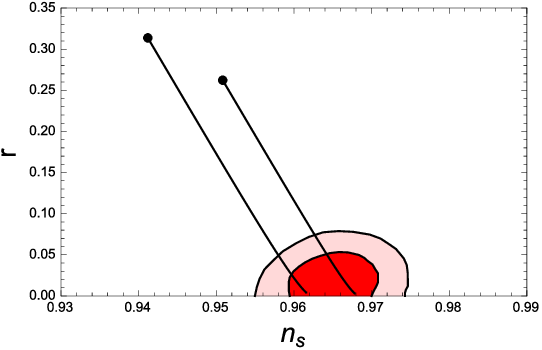}
   \end{center}
\caption{
The inflationary predictions ($n_s$ and $r$) in $\mu$-hybrid inflation 
  for various values of $\xi \geq 0$ for $N_0=50$ (left solid line) and $60$ (right solid line),   
  along with the contours for the limits at the confidence levels of $68\%$ (inner) and $95\%$ (outer) 
  obtained by the Planck 2018 measurements ({\it Planck} TT, TE, EE+lowE+lensing+BKP14)~\cite{ref7}.   
The black points correspond to the predictions of the minimal $\lambda \varphi^4$ inflation ($ \xi=0$). 
The predicted $r$ value approaches its asymptotic value 
 $r \simeq 0.00419$ for $N_0=50$ ($r \simeq 0.00296$ for $N_0=60$) 
 as $\xi$ is increased.     
}
 \label{fig:1}
\end{figure}

After imposing the constraint on the amplitude of the curvature perturbation 
\begin{eqnarray}
\Delta_R ^2 = 2.099\times10^{-9} 
\end{eqnarray}
  from the Planck measurements \cite{ref7} with the pivot scale chosen at $k_0 = 0.002$ Mpc$^{-1}$ 
  and assuming the number of e-foldings ($N_0$) to be 50 or 60 as representative values, 
  we obtain the inflationary predictions as a function of $\xi$. 
The result is shown in Figure \ref{fig:1}, along with the Planck results. 
The inflationary predictions for various $\xi$ values are summarized in Table \ref{tab:1}. 
To satisfy the limit obtained by the Planck at $95\%$ confidence level in Figure \ref{fig:1}, 
  we find a lower bound on $\xi=0.0136$ ($\xi \geq 0.00526$), 
  corresponding to $ r \leq 0.0496$ ($r \leq 0.0963$) 
  for $N_0=50$ ($N_0=60$). 
This, in turn, yields a lower bound on $\kappa$, namely
\begin{equation}
\kappa \geq  2.54 \times 10^{-6} \; (1.41\times10^{-6}).
\end{equation}
With $r \gtrsim 0.004$ ($0.003$) for $N_0=50$ ($60$), primordial gravity waves according to this scenario lie in the observable range. 
This is in sharp contrast to standard hybrid inflation which yields $r \lesssim 10^{-4}$.

\begin{table}[th]
\begin{center}
\begin{tabular}{|c||cc|ccc|c|}
\hline
\multicolumn{7}{|c|}{$N_0=50$}   \\
\hline 
 $\xi $ &  $\varphi_0/M_P$ & $\varphi_e/M_P $ & $n_s$   &  $r$  &  $\alpha (10^{-4})$  & $\kappa$\\ 
\hline
  $0$             & $20.2$  & $2.83$    &  $0.941$ &  $ 0.314$  & $-11.5 $             &  $ 9.68  \times 10^{-7}  $    \\
  $0.00527$  & $20.0$  & $2.77$    & $0.955$  &  $ 0.1$     & $-9.74$              &  $ 1.73  \times 10^{-6}  $    \\
  $0.119$    & $15.8$    & $2.07$    & $0.961$  &  $ 0.01$   & $-7.70$              &  $ 8.66  \times 10^{-6}  $    \\
  $1$             & $7.82$      & $1.00$       & $0.961$  &  $0.00489$    & $-7.51$ &  $ 5.01  \times 10^{-5}  $    \\
  $10$           & $2.65$      & $0.337$   & $0.962$  &  $0.00426$  & $-7.49$     &  $ 4.67  \times 10^{-4}    $    \\
  $1000$     & $0.267$ & $0.0340$ & $0.962$  &  $0.00419$ & $-7.48$      &  $ 4.63  \times 10^{-2}    $    \\
\hline
\hline
\multicolumn{7}{|c|}{$N_0=60$}   \\
\hline 
 $\xi $ & $\varphi_0/M_P$ & $\varphi_e/M_P $ & $n_s$   &  $r$  &  $\alpha (10^{-4})$  & $\kappa$\\ 
\hline
  $0$             & $22.1$  & $2.83$    &  $0.951$ &  $ 0.262$  & $-8.06 $           &  $ 7.40  \times 10^{-7}  $    \\
  $0.00333$  & $22.00$  & $2.79$    & $0.961$  &  $ 0.1$     & $-7.03$           &  $ 1.20  \times 10^{-6}  $    \\
  $0.0690$    & $18.9$    & $2.30$    & $0.967$  &  $ 0.01$   & $-5.44$            &  $ 5.06  \times 10^{-6}  $    \\
  $1$             & $8.52$      & $1.00$       & $0.968$  &  $0.00346$    & $-5.25$ &  $ 4.20  \times 10^{-5}  $    \\
  $10$           & $2.89$      & $0.337$   & $0.968$  &  $0.00301$  & $-5.24$     &  $ 3.92  \times 10^{-4}    $    \\
  $1000$     & $0.291$ & $0.0340$ & $0.968$  &  $0.00296$ & $-5.23$           &  $ 3.88  \times 10^{-2}    $    \\
\hline
\end{tabular}
\end{center}
\caption{ 
Inflationary predictions (scalar spectral index $n_s$, 
  tensor to scalar ratio $r$, and running of the spectral index $\alpha$)  
  for various values of $\xi$ in $\lambda \varphi^4$ inflation 
  with non-minimal gravitational coupling. 
Here $\varphi_0$ and $\varphi_e$, respectively, denote the inflaton field values at the horizon exit and 
  the end of inflation set by the condition for the slow-roll parameter $\epsilon(\varphi_e)=1$. 
} 
\label{tab:1}
\end{table}

In order to discuss inflaton decay and subsequent reheating, 
  we recall that the VEV of $S$ in the presence of the supersymmetry breaking ($\langle F_\phi \rangle = m_{3/2}$) 
  is given by $\langle S \rangle \simeq m_{3/2}/ \kappa $, for $m_{3/2} \ll M$ \cite{ref9}. 
In turn, this generates the MSSM $\mu$ term:
\begin{equation}
\mu = \frac{\sigma}{\kappa}m_{3/2}. 
\label{mu-term}
\end{equation}
The inflaton field $\varphi$ can decay, in the exact SUSY limit, into the MSSM Higgs fields 
  via the $F$-term ($F_S$) of the superfield $S$ between the superpotential terms 
  $\kappa S \bar{X} X$ and $\sigma S H_uH_d$ \cite{PRD1602.07866}. 
The decay width is given by
\begin{equation}
\Gamma = \frac{\sigma^2}{8 \pi} m_{\varphi} =  \frac{\sqrt{2}}{8\pi}\kappa^3 \left(\frac{\mu}{m_{3/2}}\right)^2 M, 
\end{equation}
where $m_{\varphi} = \sqrt{2}\kappa M$ is the inflaton mass at the potential minimum, 
  and we have assumed, for simplicity, that the inflation can decay into all the Higgs bosons in the MSSM.    
We estimate the reheat temperature $T_r$ from the relation,
\begin{equation}
\Gamma = H(T_r) =\left(\frac{\pi^2}{90}g_*\right)^{1/2}\frac{T_r^2}{M_P}.
\end{equation}
Setting $g_* \simeq 200$ for the MSSM, we find in units of GeV 
\begin{eqnarray}
T_r  \simeq 1.71 \times10^8 \;  \kappa^{3/2}\left(\frac{\mu}{m_{3/2}}\right) M^{1/2}.  
\label{T_r}
\end{eqnarray}

\begin{figure}[t]
  \begin{center}
   \includegraphics[width=8cm]{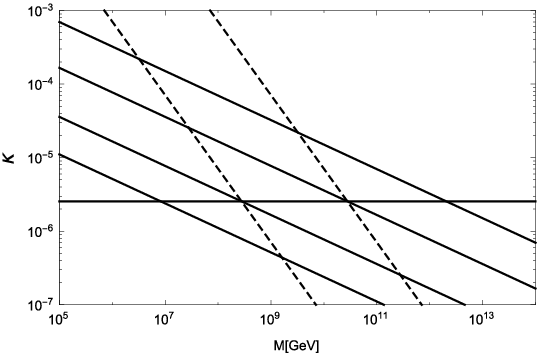}
   \end{center}
\caption{
Allowed parameter regions in ($M$, $\kappa$)-plane. 
The diagonal solid lines correspond to the upper bounds on $\kappa$ from the gravitino BBN constraint 
  of Eq.~(\ref{BBN}) for various values of $\mu/m_{3/2}=1$, $8.6$, $86$ and $500$, respectively,  
  from top to bottom.  
The dashed diagonal lines corresponds to the lower bound on $\kappa$ of Eq.~(\ref{decay}) 
 with $T_r^{\rm max}=10^6$ GeV for $M_H=1$ TeV and $M_H=100$ TeV, respectively, from left to right.  
The horizontal solid line depicts the lower bound on $\kappa$ from the Planck measurements, 
  $\kappa > 2.54 \times 10^{-6}$ for $N_0=50$.  
For a fixed  $\mu/m_{3/2}$ value, a region surrounded by three lines 
  (below the solid diagonal line and above the dashed and horizontal solid lines) 
  is allowed.  
For $\mu/m_{3/2} > 86$, no allowed region exists for $M_H=1$ TeV. 
If we take a larger value for $T_r^{\rm max}$, the diagonal solid lines shift upward (see Eq.~ Eq.~(\ref{decay})) 
  and allowed regions are enlarged. 
}
 \label{fig:2}
\end{figure}

Here, we comment on a justification of our estimate for the reheat temperature in Eq.~(\ref{T_r}). 
In $\lambda \varphi^4$ inflation with non-minimal gravitational coupling, 
   in particular the Higgs inflation,   
   the reheat temperature is expected to be very high, $T_r \gtrsim 10^{13}$ GeV, 
   because of the parametric resonance effects \cite{HiggsInf, H-S}.  
However, we notice that the parametric resonance effects discussed in Refs.~\cite{HiggsInf, H-S} 
   are important for a limited case,  namely, $\xi \gg 1$. 
From Figure \ref{fig:2}, we have an upper bound on $\kappa \lesssim10^{-4}$, 
   which means $\xi$ is not large: $\xi \lesssim 10$ from Table \ref{tab:1}.   
As previously studied in, for example, Ref.~\cite{Preheat}, 
   even if a non-minimal gravitational coupling is zero,  
   an inflaton oscillating in its quartic potential can give rise to resonant production  
   of a particle (Higgs bosons in our case) through its coupling. 
To see if the parametric resonance effect is important in our model,  
   we follow the analysis in Ref.~\cite{Ballesteros:2016xej}.    
From our superpotential, we have a coupling between the inflaton and the Higgs doublets 
  in the scalar potential given by
\bea
   V_{SUSY} \supset \frac{1}{4} \kappa \sigma \varphi^2 (H_u H_d + h.c.).  
\eea
For inflaton values $|\varphi | \gg M$ during its oscillation in the quartic potential $\kappa^2 \varphi^4/16$, 
  the Higgs doublets acquire an effective mass $m_H \simeq  \sqrt{\kappa \sigma} | \varphi |$. 
This effective mass is to be compared with the effective frequency of the oscillating inflaton, $\omega \simeq \kappa  | \varphi |$. 
Since we take $\sigma > \kappa$, the effective mass is larger than the frequency (for $| \varphi | \gg M$), 
  so that the resonant production of Higgs doublets is not effective. 
When $m_H$ becomes smaller than $\omega$, the resonant production may begin. 
However, the condition $m_H < \omega$ is satisfied when the inflaton gets close to the potential minimum, 
 $| \varphi | \simeq M$. 
In this case, the inflaton is oscillating around the potential minimum with $\omega \simeq m_\varphi$, 
  and hence, our estimate of the reheat temperature by the inflation decay width is reasonable, 
  neglecting the parametric resonance effects as discussed above.

\begin{figure}[t]
  \begin{center}
   \includegraphics[width=8cm]{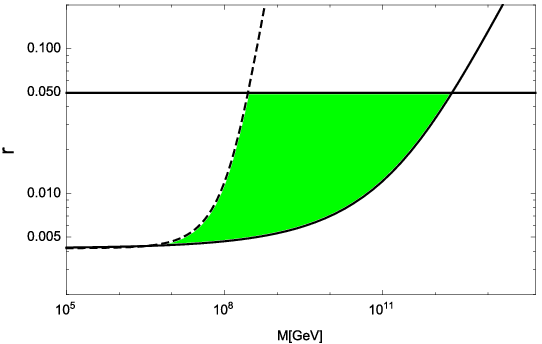}
   \end{center}
\caption{
Allowed parameter regions in ($M$, $r$)-plane. 
The solid curve corresponds to the solid diagonal line (top), 
  while the dashed curve corresponds to the (left) dashed diagonal line in Figure \ref{fig:2}. 
The horizontal solid line depicts the upper bound from the Planck measurements, 
  $r \leq 0.0496$ for $N_0=50$. 
The shaded region satisfies all the constraints. 
}
 \label{fig:3}
\end{figure}

To proceed further we first assume that the gravitino is not the LSP. 
Imposing the constraint $T_r  \leq T_r^{\rm max} \simeq 10^6-10^9$ GeV 
  for $m_{3/2}= 1- 10\, {\rm TeV}$  
  in order to avoid the cosmological gravitino problem \cite{Moroi},
we find  
\begin{equation}
\kappa \leq 0.0324 \left( \frac{T_r^{\rm max}}{10^6 \, {\rm GeV}}\right)^{2/3}\left(\frac{m_{3/2}}{\mu}\right)^{2/3} 
 \left( \frac{1 \; {\rm GeV}}{M} \right)^{1/3}. 
\label{BBN}
\end{equation}
In order to keep the inflaton decay channel open into all the Higgs bosons in the MSSM,   
  we impose a lower bound on the inflaton mass as $m_\phi \geq M_H$ and hence, 
\begin{equation}
   \kappa \geq \frac{1}{\sqrt{2}} \left( \frac{M_H}{M} \right), 
\label{decay}
\end{equation}
where $M_H$ is the mass scale of the heavy Higgs bosons in the MSSM. 
Varying $\mu/m_{3/2}$ from $1$ to $500$, a plot of $\kappa$ versus $M$ is shown in Figure \ref{fig:2}. 
Corresponding to a low rehearing temperature $T_r \leq 10^6$ GeV, we have set $N_0=50$ \cite{Lyth:1998xn}. 
Note that a solution exists only for $\mu/m_{3/2} \leq 86$ (8.6) for $M_H=1$ (100) TeV.  
In summary, for $T_r \leq 10^6$ GeV and $M_H=1$ TeV, the symmetry breaking scale $M$ 
 is constrained as follows: 
\begin{equation}
2.8\times10^8 \leq M[{\rm GeV}] \leq 2.1\times10^{12}.
\end{equation}
Note that $M$ values of order $M_{GUT}=2 \times 10^{16}$ GeV are realized 
  if we set $T_r \simeq 10^8$ GeV.\footnote{
Such a higher $T_r$ is allowed if we take $m_{3/2}\simeq 10$ TeV \cite{Moroi}.
}  
These results should be contrasted with conventional $\mu$-hybrid inflation \cite{ref1, ref9, OS},  
  which yield $M$ values of the order of $10^{15}$-$10^{16}$ GeV.
Using a one-to-one correspondence between the $\kappa$ value and the inflationary predictions, 
  we show in Figure \ref{fig:3} a plot of $r$ versus $M$ for $\mu/m_{3/2}=1$.     

\begin{figure}[t]
  \begin{center}
   \includegraphics[width=8cm]{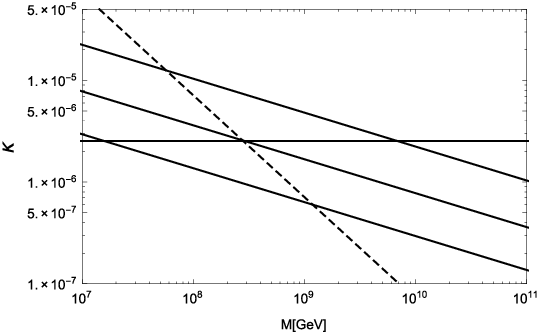}
   \end{center}
\caption{
Allowed parameter regions in ($M$, $\kappa$)-plane. 
The diagonal solid lines correspond to $m_{3/2}=3.2$, $1.45$ and $0.7$ GeV, respectively,  
  from top to bottom, along which the observed dark matter relic density is reproduced 
  (see Eq.~(\ref{DMrelic})). 
Here, we have taken $\mu=1$ TeV and $M_3=3$ TeV, for simplicity.    
The dashed diagonal line corresponds to the lower bound on $\kappa$ of Eq.~(\ref{decay}) 
  for $M_H=1$ TeV. 
The horizontal solid line depicts the lower bound on $\kappa$ from the Planck measurements,  
  $\kappa \geq 2.54 \times 10^{-6}$, corresponding to $r \leq 0.0496$ for $N_0=50$.  
For a fixed  gravitino mass, a region surrounded by three lines 
  (below the solid diagonal line and above the dashed and horizontal solid lines) 
  is allowed.  
For $m_{3/2} < 1.45$ GeV, no allowed region exists. 
}
 \label{fig:4}
\end{figure}

Next let us explore the gravitino LSP scenario. 
The gravitino relic density from thermal production is given by \cite{gravitinoDM}
\begin{equation}
\Omega h^2 \simeq 1.8\times10^{-8} \left(\frac{T_r}{m_{3/2}}\right)\times\left(\frac{M_3}{3 \; {\rm TeV}}\right)^2,
\end{equation}
where $M_3$ is a running gluino mass. 
In oder to reproduce the observed dark matter relic abundance, $\Omega h^2\simeq 0.1$ \cite{Planck-2}, we find 
\begin{eqnarray}
%
 \kappa = 0.102\left(\frac{m_{3/2}[{\rm GeV}]^2}{\mu[{\rm GeV}] \sqrt{M{\rm [GeV]}}}\right)^{2/3}\left(\frac{3\ {\rm TeV}}{M_3}\right)^{4/3}.  
\label{DMrelic}
\end{eqnarray}
In the gravitino LSP scenario, we also have the BBN constraint on the next LSP (NLSP) lifetime given by \cite{NLSP}
\begin{equation}
\tau_{NLSP} \simeq 10^4 \; {\rm sec} \left(\frac{100 \ {\rm GeV}}{m_{NLSP}}\right)^5 \left(\frac{m_{3/2}}{1 \ {\rm GeV}}\right)^2 < 1 \; {\rm sec} .
\end{equation}


As an example, let us consider $\mu \sim 1 \; {\rm TeV} \sim m_{NLSP}$, which yields 
\begin{equation}
m_{3/2} \lesssim 3.2 \; {\rm GeV} 
\end{equation}
from the BBN constraint. 
In Figure \ref{fig:4} we plot $\kappa$ versus $M$ for various $m_{3/2}$ values. 
A gravitino dark matter scenario is viable with $m_{3/2} \gtrsim 1.45$ GeV.

A discussion about inflation is not complete without at least mentioning
how the observed baryon asymmetry is realized. In our case this can be
achieved via thermal \cite{ref14} or non- thermal \cite{ref15} leptogenesis by introducing right handed neutrinos. For instance,
identifying U(1) with a local $U(1)_{B-L}$ symmetry requires three right
handed neutrinos, one per family, to avoid gauge anomalies, which have Yukawa couplings with the
inflaton system. With at least one of the right handed neutrinos
lighter than the inflaton, the decay products of the latter would
include this neutrino and leptogenesis becomes possible.

In summary, the inflationary scenario we have described can be based on a fairly minimal, 
 namely, a local U(1) extension of the MSSM or a more elaborate one such as grand unification. 
If the gravitino is not the LSP, the BBN constraint leads to a lower bound on the tensor to scalar ratio 
  of $r \gtrsim  0.004$ and the U(1) symmetry breaking scale lies in the range of  $10^8$-$10^{12}$ GeV. 
A LSP gravitino with mass of order 1 GeV is a viable dark matter candidate.


\section*{Acknowledgments}
N.O. would like to thank the Particle Theory Group of the University of Delaware for hospitality during his visit. 
This work is supported in part by the DOE Grant No.~DE-SC0013680 (N.O.) and No.~DE-SC0013880 (Q.S.).


\end{document}